\documentclass[
  aps,
  prb,
  reprint,
  superscriptaddress,
  longbibliography,
  nofootinbib
]{revtex4-2}

\usepackage{amsmath,amssymb}
\usepackage{graphicx}
\usepackage{bm}
\usepackage{hyperref}
\hypersetup{colorlinks=true,linkcolor=blue,citecolor=blue,urlcolor=blue}

\begin{document}

\title{Static vacancies as parametrized conformal defects in the
       critical $J_1$--$J_2$ transverse-field Ising chain}

\author{R. L. Silva}
\email[Corresponding author: ]{ricardo.l.silva@ufes.br}
\affiliation{Departamento de Ci\^encias Naturais, Universidade Federal do Esp\'irito Santo, S\~ao Mateus, ES 29932-540, Brazil}

\author{R. C. Silva}
\affiliation{Departamento de Ci\^encias Naturais, Universidade Federal do Esp\'irito Santo, S\~ao Mateus, ES 29932-540, Brazil}

\author{R. J. C. Lopes}
\affiliation{Departamento de F\'isica, Universidade Federal do Rio Grande do Norte, Natal, RN, Brazil}

\author{A. R. Pereira}
\affiliation{Departamento de F\'isica, Universidade Federal de Vi\c{c}osa, Vi\c{c}osa, 36570-000 Minas Gerais, Brazil}

\date{\today}

\begin{abstract}
We revisit the problem of two static nonmagnetic vacancies in the
transverse-field Ising chain with first- and second-neighbor couplings
$J_1$ and $J_2$, now  on the critical line, using
density-matrix renormalization-group (DMRG) calculations in open
chains of up to $N=300$ sites. In contrast to the gapped regime
studied previously [Silva {\it et al.}\cite{SilvaGuimaraesPereira2005}], where the vacancy-vacancy interaction decays
exponentially, along the entire quantum critical line the interaction
becomes algebraic, $|\Delta_b(r)|\sim r^{-\alpha}$, with $\alpha$
close to the universal Casimir value of unity and a weak but
systematic dependence on the second-neighbor coupling,
$\alpha_\infty \simeq 1.070 + 0.091\,(J_2/J_1)$ across
$J_2/J_1\in[0.1,1.0]$. The transmission ratio of the spin correlator 
across a vacancy approaches a $J_2$-dependent plateau 
$T_\infty(J_2)$ that grows from $0.11$ to $0.33$ over the same range,
and the Affleck-Ludwig boundary entropy is small and approximately
constant, $\log g_\infty \approx -0.073$, well above the Ising
fixed-BC value $-\ln\sqrt{2}$ and close to the free-boundary value.
The three observables vary smoothly and monotonically with $J_2$,
consistent with a one-parameter family of partially transmissive 
conformal defects controlled by $J_2$.
Throughout, the critical line is located using the bulk
spin-correlator exponent $\eta=1/4$, the order-parameter exponent of
the Ising universality class, which provides a robust criterion
in this open geometry.
\end{abstract}

\keywords{quantum phase transitions, conformal field theory, spin chains, entanglement in extended quantum systems}

\maketitle

\section{Introduction}

The interplay between impurities and quantum criticality is a
long-standing problem in condensed matter physics. In one dimension,
a nonmagnetic impurity inserted into a critical spin chain behaves as a
conformal defect: a localized object that preserves the bulk
conformal symmetry while reflecting and transmitting low-energy
excitations, characterized by a small set of universal
numbers~\cite{AffleckLudwig1991,eggertAffleck1992,OshikawaAffleck1996,OshikawaAffleck1997,
FriedanKonechny2004,CalabreseCardy2004,Calabrese2009}.
The boundary degeneracy $\log g$~\cite{AffleckLudwig1991}, the leading
boundary correction to the entanglement
entropy~\cite{CalabreseCardy2004}, and the transmission of bulk
operators~\cite{RoySaleur2022,Bachas2002,QuellaRunkelWatts2007}
provide complementary windows on the physics of defects.

In an earlier work~\cite{SilvaGuimaraesPereira2005}, we studied two
static vacancies in the ferromagnetic transverse-field Ising chain
with first- and second-neighbor exchange,
\begin{equation}
H \;=\; -J_1 \!\sum_i \sigma^x_i \sigma^x_{i+1}
        -J_2 \!\sum_i \sigma^x_i \sigma^x_{i+2}
        -b   \!\sum_i \sigma^z_i ,
\label{eq:H}
\end{equation}
and found, by exact diagonalization in small chains and in the 
gapped (ferromagnetic) phase below the critical field, that the
field-induced effective interaction between the vacancies decays
exponentially with their separation. This behavior is expected because in that regime the bulk correlation length is finite, the vacancy disturbance is localized, and the resulting Casimir interaction~\cite{krech1994} is screened.

The situation along the quantum critical line of
Eq.~\eqref{eq:H} is qualitatively different. The bulk correlation
length diverges, screening is absent, and the vacancies--each acting
as a boundary perturbation--are expected to interact through power-law Casimir forces of conformal-field-theoretic origin. The exponent and
the prefactor of this interaction encode the universal data of the
defect~\cite{PhysRevLett.99.185301}.

In this work, we extract these data using DMRG in open chains of up
to $N=300$ sites, working directly on the critical line of
Eq.~\eqref{eq:H}. The key observables are the binding energy
$\Delta_b(r)$ as a function of vacancy separation, the transmission
ratio of the spin correlator to a single vacancy, and the
boundary entropy $\log g$. Our central findings are:
(i) the binding energy decays as a clean power law,
$|\Delta_b(r)| \sim r^{-\alpha}$, with $\alpha$ close to unity and a
weak systematic increase with $J_2$, across the entire critical line;
(ii) $\alpha$, the asymptotic transmission amplitude $T_\infty$, and
$\log g$ all vary smoothly and monotonically with $J_2$, consistent
with a family of one-parameter conformal defects controlled by $J_2$;
and (iii) the critical line can be located throughout the bulk
order-parameter exponent $\eta=1/4$, which we find to be the most
robust diagnostic of Ising criticality in this open geometry.

The paper is organized as follows. Section~\ref{sec:model} defines
the model and the vacancy geometries. Section~\ref{sec:method}
describes the implementation of the DMRG and the calibration of the critical
field. Sections~\ref{sec:alpha}, \ref{sec:transmission}, and
\ref{sec:logg} present the results for the Casimir exponent $\alpha$,
the transmission $T_\infty$, and the boundary entropy $\log g$,
respectively. Section~\ref{sec:discussion} discusses the unified
picture and its relation to the literature. We summarize in
Sec.~\ref{sec:conclusion}.

\section{Model and observables}
\label{sec:model}

We study the Hamiltonian of Eq.~\eqref{eq:H} on an open chain of $L$
sites. A nonmagnetic vacancy is implemented at the site $v$ by removing
the spin from that site: all $J_1$ and $J_2$ bonds touching $v$ are
simply absent, while the $J_2$ bond bridging across the vacancy
(between $v-1$ and $v+1$, at the original-lattice distance two) is
retained. This is the same vacancy prescription used in
Ref.~\cite{SilvaGuimaraesPereira2005}; it preserves the rigid-lattice
distances and allows the $J_2$ coupling to provide a tunable transmission
channel across the vacancy. Consequently---a point central to the subsequent 
analysis---the chain with a vacancy remains a
single connected system of $N-1$ spins: the vacancy is a bulk defect,
not a cut.

\begin{figure}[t]
\centering
\includegraphics[width=\columnwidth]{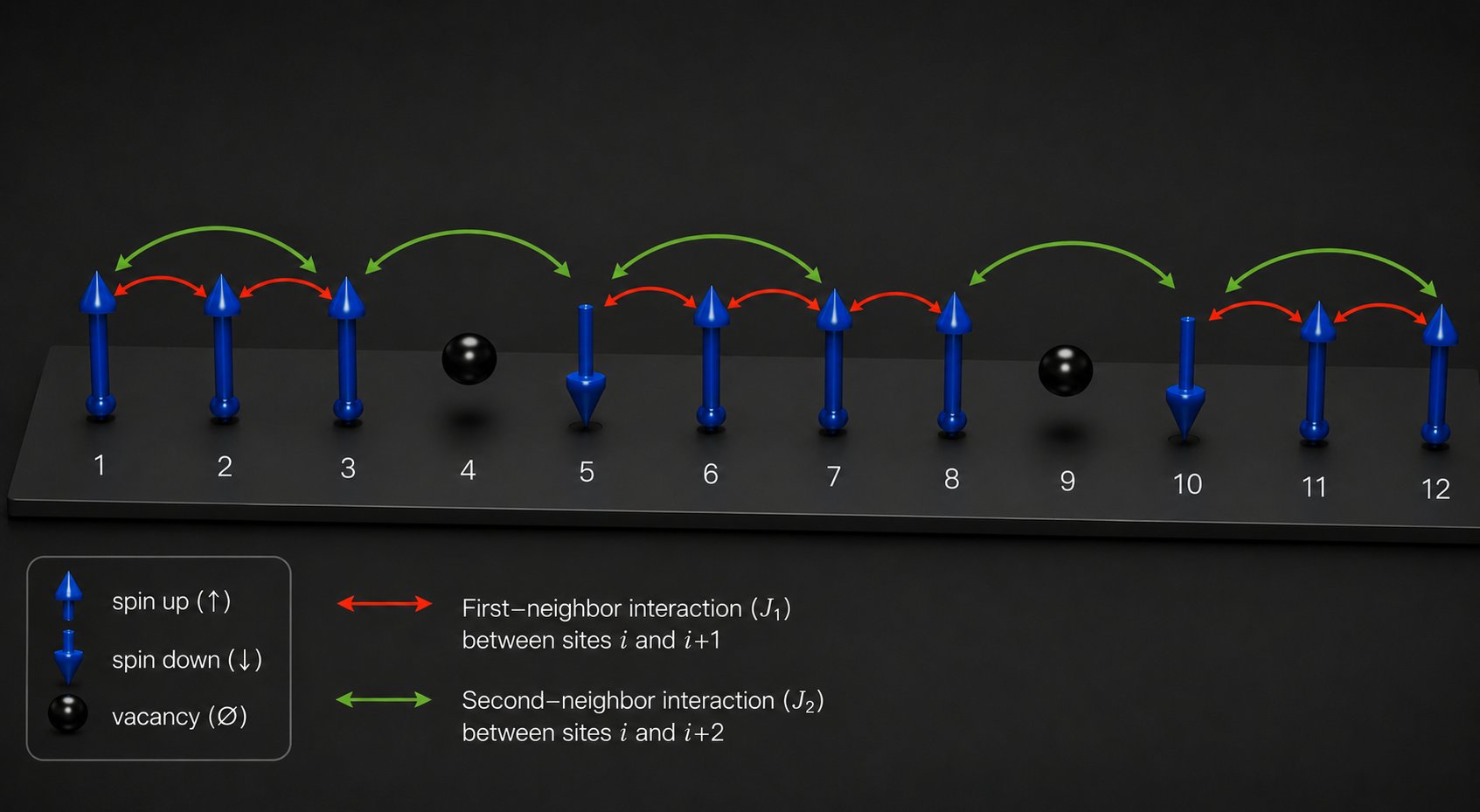}
\caption{The $J_1$--$J_2$ transverse-field Ising chain of
Eq.~\eqref{eq:H} with two static nonmagnetic vacancies a distance $r$
apart. Black spheres are vacant sites; the remaining spins interact
via $J_1$ (red) and $J_2$ (green). The $J_2$ bond bridging across
each vacancy (orange) keeps the chain connected and is the channel
that gives the defect its $J_2$-dependent transmission.}
\label{fig:model}
\end{figure}

We work on the quantum critical line $b=b_c(J_2)$ of the clean chain
(see Sec.~\ref{sec:method} for how $b_c$ is determined). In this
regime the bulk is in the Ising universality class for all $J_2/J_1$
studied here, with central charge $c=1/2$ and bulk spin-correlator
exponent $\eta=1/4$; at $J_2=0$ the clean chain reduces to the exactly
solvable transverse-field Ising model~\cite{Pfeuty1970}.

We extract three observables:

\paragraph*{Binding energy.}
Given the ground-state energies $E_0$ (clean chain), $E_{1v}$ (one
central vacancy), and $E_{2v}(r)$ (two vacancies a distance $r$ apart,
centered about the chain midpoint), we define
\begin{equation}
\Delta_b(r) \;=\; E_{2v}(r) - 2\,E_{1v} + E_0 ,
\label{eq:Db}
\end{equation}
the same definition as in Ref.~\cite{SilvaGuimaraesPereira2005}. This
is the Casimir-like interaction energy between the two vacancies,
free of self-energies. We fit $|\Delta_b(r)|\sim r^{-\alpha}$ in the
window $r\in[4,32]$ and extrapolate the exponent in $1/N$ to obtain
$\alpha_\infty(J_2)$.

\paragraph*{Defect transmission.}
For the single-vacancy geometry, with $v=L/2$, we compute
$\langle \sigma^x_i \sigma^x_j \rangle_{\rm vac}$ from the DMRG state
and form the ratio
\begin{equation}
T(k) \;=\; \frac{|\langle \sigma^x_{v-k}\, \sigma^x_{v+k}\rangle_{\rm vac}|}
                 {|\langle \sigma^x_{v-k}\, \sigma^x_{v+k}\rangle_{\rm clean}|} ,
\label{eq:T}
\end{equation}
which probes the across-defect transmission~\cite{RoySaleur2022,Bachas2002,QuellaRunkelWatts2007}.
In all cases $T(k)$ approaches a finite plateau $T_\infty$ as $k$
increases, characterizing a partially transmissive conformal defect
(rather than a power-law decay).

\paragraph*{Boundary entropy.}
From the von Neumann entanglement entropy $S(\ell)$ of the bipartition
at bond $\ell$ of the open chain~\cite{VidalLatorreRicoKitaev2003},
the Affleck-Ludwig boundary
entropy~\cite{AffleckLudwig1991,CalabreseCardy2004} appears as the
subleading constant in the Calabrese-Cardy form
$S(\ell)= (c/6)\ln\!\left[(2N/\pi)\sin(\pi\ell/N)\right] + \log g + {\rm const}$.
Because the vacancy chain is a single connected system of $N-1$
sites, the defect contribution must be isolated by subtracting the
clean-chain entropy at matched conformal coordinates: a bond at
the original-lattice index $\ell$ in the vacancy chain corresponds to
reduced position $\ell_{\rm red}=\ell$ for $\ell<v$ and
$\ell_{\rm red}=\ell-1$ for $\ell>v$, in a chain of effective length
$N-1$. Forming
$\Delta S(\ell) = S_{\rm vac}(\ell_{\rm red};N-1) - S_{\rm clean}(\ell;N)$
with the bulk logarithm evaluated at the appropriate length for each
term, the bulk contribution and the shared free-end constant cancel,
and the residual is governed by the defect alone. (A naive
subtraction at equal original indices, ignoring the unit coordinate
shift and the $N\to N-1$ length change, instead produces a spurious
cusp centered on the vacancy and an unstable estimate; see
Appendix~\ref{app:data}.) We summarize the residual profile by its
median over a bulk window away from the defect and the chain ends, as
detailed in Sec.~\ref{sec:logg}.

\section{Method: DMRG and calibration of \texorpdfstring{$b_c$}{bc}}
\label{sec:method}

\begin{figure*}[t]
\centering
\includegraphics[width=0.92\textwidth]{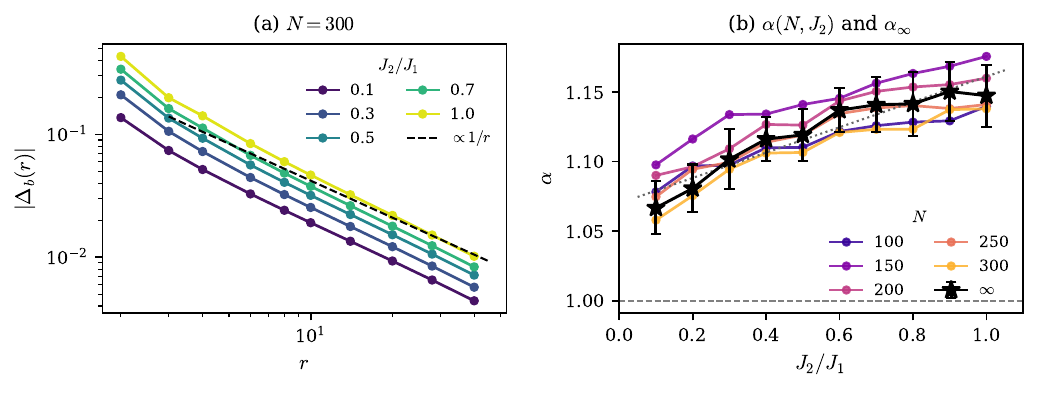}
\caption{(a) The vacancy-vacancy binding energy $|\Delta_b(r)|$ on the
critical line, $N=300$, for representative $J_2/J_1$ values. The
behavior is power-law to high precision (linear in log-log over more
than a decade in $r$); $1/r$ is shown for reference. The fits in the
window $r\in[4,32]$ all have $R^2\geq 0.996$.
(b) The fitted exponent $\alpha(N,J_2)$ for the five sizes (colored
symbols) and the $1/N$-extrapolated $\alpha_\infty(J_2)$ (black stars,
error bars from the standard error of the linear extrapolation). The
dotted line is the linear fit of Eq.~\eqref{eq:alpha_fit}; the dashed
line marks $\alpha=1$.}
\label{fig:alpha}
\end{figure*}

We use a standard two-site DMRG~\cite{White1992,Schollwock2005,SCHOLLWOCK201196} based
on ITensor~\cite{ITensor} for the Hamiltonian of Eq.~\eqref{eq:H} on
open chains of $L=N \in \{100,150,200,250,300\}$ sites. To avoid convergence to low-entanglement local minima, which can plague critical
DMRG started from a paramagnetic product state, we use a N\'eel-type
initial state $|\!\uparrow\downarrow\uparrow\downarrow\cdots\rangle$
and a strong, slowly decaying mixer (noise of amplitude $10^{-2}$
sustained over 20 sweeps before decay). The 
bond dimension is allowed
to grow to $\chi=600$; in practice, the converged critical-point
ground states require modest 
bond dimensions 
that grow slowly with
the size of the system, with the median $\chi\simeq 60$ and the maximum $\chi=91$ at
$N=300$ and the truncation error $\sim 10^{-10}$.

\subsection{Calibration of the critical field}

We locate the critical line using the bulk spin--spin correlator. At
the Ising critical point, the order-parameter correlator decays as
$\langle\sigma^x_i\sigma^x_j\rangle \sim |i-j|^{-\eta}$ with
$\eta=1/4$; measured in the central region of the open chain, away
from both boundaries, this exponent is a sharp and direct locator of
criticality. Specifically, we evaluate the correlator on pairs of
sites placed symmetrically about the chain center, with separations
$4\le r \le 2N/5$, and fit a single power law to this window. For
each $(N,J_2)$ we then determine $b_c$ by bisection until
$|\eta(b_c)-0.25|<0.005$, requiring $R^2>0.97$ for the correlator fit.
This procedure yields $\eta=0.250\pm 0.003$ at all 50 calibrated
points, with mean fit quality $R^2=0.992$. At these system sizes, the effective exponent extracted from an open chain retains a residual dependence on the choice of fit window, at the level of a few percent
of $\eta$; we therefore use the identical window at every $(N,J_2)$,
so that the calibration is internally consistent across the whole
data set, and we verify that $b_c(N)$ converges smoothly with $N$
(the change between $N=250$ and $N=300$ is at most $2.5\times10^{-3}$
over the entire range of $J_2$). That $\eta=1/4$ holds throughout
$J_2/J_1\in[0.1,1.0]$ confirms that the bulk remains in the Ising
universality class along the entire critical line, consistent with
the second-neighbor $J_2$ term being irrelevant at the Ising fixed
point.

We do not use the entanglement-entropy central charge as a calibration
criterion. In this open geometry, a naive Calabrese--Cardy fit of
$S(\ell)$ over a central window of cuts is known to overestimate the
slope~\cite{laflorencie20161}, because of the alternating (parity) correction and the
curvature of the conformal-distance logarithm near $\ell=N/2$; we have
checked, on the exactly solvable clean chain at $J_2=0$ using the
free-fermion correlation-matrix
method~\cite{Peschel2003,PeschelEisler2009}, that this fit returns an
apparent $c\simeq0.75$ even though the true value is $c=1/2$, while a
half-chain estimator $S(N/2)\sim(c/6)\ln N$ recovers $c\to1/2$. The
bulk order-parameter exponent $\eta$, measured from the central spin
correlator, is free of this boundary artifact and provides a sharper,
more direct criterion for locating the critical field; we therefore
rely on $\eta=1/4$ throughout. This boundary bias does not affect the
boundary entropy of Sec.~\ref{sec:logg}, which is obtained by a
matched subtraction that cancels the bulk logarithm.

\section{Casimir interaction: exponent \texorpdfstring{$\alpha$}{alpha}}
\label{sec:alpha}

Figure~\ref{fig:alpha}(a) shows $|\Delta_b(r)|$ for $N=300$ at five
representative values of $J_2/J_1$. The data exhibits a clear linear behaviour in log-log over more than a decade in $r$, demonstrating that
the vacancy-vacancy interaction along the critical line is genuinely
algebraic, in sharp contrast to the exponential decay found in the
gapped regime in Ref.~\cite{SilvaGuimaraesPereira2005}. This is the
qualitative change in the vacancy physics induced by quantum criticality.

A power-law fit in the window $r\in[4,32]$ gives, for each $(N,J_2)$,
an effective exponent $\alpha(N,J_2)$ with $R^2$ ranging from $0.996$
to $1.000$. The values are shown in Fig.~\ref{fig:alpha}(b). At fixed
$J_2$, $\alpha$ varies by at most $\approx 0.04$ between $N=100$ and
$N=300$; the dependence on $1/N$ is mild but not strictly monotonic,
which limits the precision of the extrapolation. We obtain
$\alpha_\infty(J_2)$ by linear extrapolation in $1/N$; the standard
error of the extrapolated intercept is about $\pm0.02$ in each $J_2$.

The extrapolated $\alpha_\infty(J_2)$, also shown in
Fig.~\ref{fig:alpha}(b), increases approximately linearly with $J_2$:
\begin{equation}
\alpha_\infty(J_2) \;\simeq\; 1.070 + 0.091\,(J_2/J_1),
\label{eq:alpha_fit}
\end{equation}
varying from $\alpha_\infty\simeq1.07$ for $J_2/J_1=0.1$ to
$\alpha_\infty\simeq1.15$ for $J_2/J_1=1.0$, with mean
$\langle \alpha_\infty\rangle = 1.12 \pm 0.03$. The exponent is thus
close to the universal Casimir value $\alpha=1$, lying slightly above
it across the studied range. Given the limited fit window and the
non-monotonic finite-size dependence, we regard the robust feature of
these data as the systematic increase of $\alpha$ with $J_2$
(slope $\simeq 0.09$), rather than the precise size of the offset
above unity, which we do not claim to resolve beyond the finite-size
scatter. In particular, the largest size ($N=300$) consistently yields the
smallest $\alpha$ at each $J_2$, so the asymptotic exponent may be
somewhat closer to unity than the $1/N$ extrapolation suggests.

\section{Defect transmission: plateau \texorpdfstring{$T_\infty(J_2)$}{T(J2)}}
\label{sec:transmission}

\begin{figure*}[t]
\centering
\includegraphics[width=0.92\textwidth]{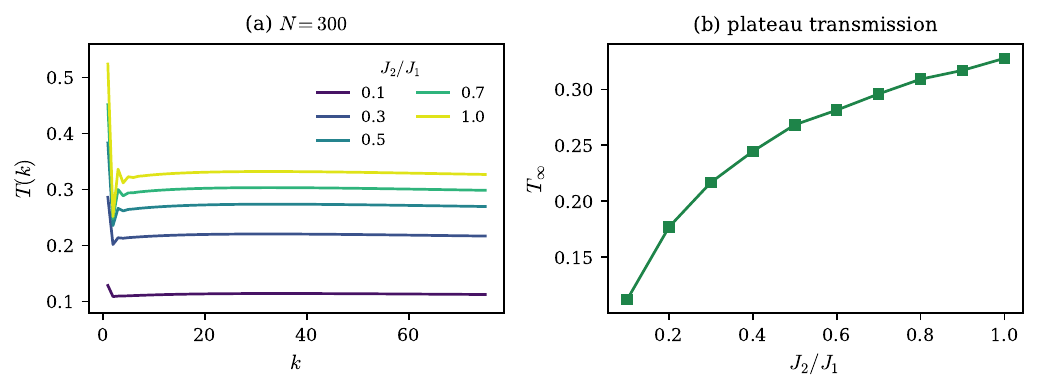}
\caption{(a) Transmission ratio $T(k)$ across a single central
vacancy for $N=300$ in some representative $J_2/J_1$ values. After a short
transient at small $k$, $T(k)$ saturates to a $J_2$-dependent plateau.
(b) The $1/N$-extrapolated plateau value $T_\infty(J_2)$ (error bars
from the extrapolation are smaller than the symbols). The defect is
strongly reflective at small $J_2$ ($T_\infty\simeq 0.11$) and becomes
substantially more transmissive at $J_2/J_1=1.0$
($T_\infty\simeq 0.33$).}
\label{fig:transmission}
\end{figure*}

The transmission ratio of Eq.~\eqref{eq:T} is shown in
Fig.~\ref{fig:transmission}(a) for $N=300$. After a short transient at
$k\lesssim 8$, $T(k)$ reaches a plateau that remains essentially
constant out to the largest distances accessible (limited by proximity
to the chain boundaries). The plateau value, calculated as the median
of $T(k)$ in the window $k\in[N/8,N/4]$ and extrapolated to
$N\to\infty$ in $1/N$, strongly depends on $J_2/J_1$:
as shown in Fig.~\ref{fig:transmission}(b), $T_\infty$ rises
monotonically from $0.112$ in $J_2/J_1=0.1$ to $0.328$ in
$J_2/J_1=1.0$. The finite-size dependence of the plateau is very
weak, so the extrapolation uncertainties are at the $10^{-3}$ level.

That $T(k)$ saturates to a finite plateau rather than decaying as a
power-law is consistent with the vacancy behaving as a partially transmissive conformal
defect: it neither perfectly transmits ($T_\infty=1$, trivial defect)
nor perfectly reflects ($T_\infty\to 0$ at large $k$). The amplitude
$T_\infty$ characterizes the strength of the defect, and $J_2$ tunes it continuously.

\section{Boundary entropy \texorpdfstring{$\log g$}{log g}}
\label{sec:logg}

\begin{figure}[b]
\centering
\includegraphics[width=0.95\columnwidth]{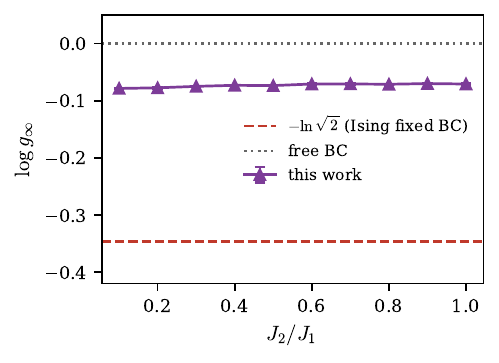}
\caption{Affleck-Ludwig boundary entropy $\log g_\infty$ of the
vacancy defect, extracted by matched-coordinate entropy subtraction
(effective length $N-1$, reduced bond position, $c=1/2$) and $1/N$
extrapolation. Error bars show the extrapolation standard error; an
additional method systematic of order $0.005$ from the choice of fit
window is discussed in the text. The dashed line marks the Ising
fixed-BC value $-\ln\sqrt 2 \approx -0.347$; the dotted line at zero
corresponds to free BC. The vacancy defect sits much closer to free
BC than to fixed BC, $\log g_\infty \approx -0.073$.}
\label{fig:logg}
\end{figure}

Figure~\ref{fig:logg} shows the boundary entropy $1/N$-extrapolated 
$\log g_\infty(J_2)$. For each $(N,J_2)$ we subtract the bulk
Calabrese-Cardy logarithm (with $c=1/2$ fixed, which removes the
open-boundary slope bias discussed in Sec.~\ref{sec:method}) from
$S_{\rm vac}$ at matched conformal coordinates, using effective
length $N-1$ and the reduced bond position; the result is referenced
to the clean-chain constant and summarized by its median over a bulk
window $0.2(N-1)\le \ell_{\rm red}\le 0.8(N-1)$, excluding bonds
within $0.06\,N$ sites of the defect.

We highlight two key properties of this estimate. First, the
matched-coordinate subtraction removes the spurious cusp produced by
a naive index-matched subtraction (Appendix~\ref{app:data}), but the
residual profile $\Delta S(\ell)$ is not strictly flat: it retains a
slow variation across the window, reflecting the crossover of the
block entanglement around a partially transmissive
defect~\cite{EislerPeschel2010}. The window median is a stable scalar
summary of this profile: it changes by less than $0.01$ between
$N=100$ and $N=300$ at fixed $J_2$, and by about $\pm0.005$ under
reasonable variations of the window boundaries. Second, with this
systematic in mind, the extrapolated values lie in the narrow range
$\log g_\infty \approx -0.073$, with a slight residual drift with
$J_2$ (from $\approx -0.079$ at small $J_2$ to $\approx -0.071$ at
$J_2/J_1=1.0$). Averaged over the studied range,
$\log g_\infty = -0.073 \pm 0.003\,({\rm stat}) \pm 0.005\,({\rm
window})$. The drift with $J_2$ is at the level of the combined
uncertainty and we do not assign physical significance to it; the
robust statement is that $\log g_\infty$ is small, negative, and
approximately constant across the entire critical line.

The defect is therefore much closer to a free open end ($\log g = 0$)
than to an Ising fixed boundary
($\log g = -\ln\sqrt 2 \approx -0.347$). The continuous variation of the entanglement of a critical Ising chain with the strength of an interface defect was established by Eisler and
Peschel~\cite{EislerPeschel2010}; here the defect is a connected,
partially transmissive vacancy and we extract its boundary entropy
directly. The qualitative conclusion--that the vacancy defect lies
close to the free-boundary value and well above the fixed-BC value
across the entire critical line--is robust and follows naturally
from the connected, partially transmissive nature of the defect: a
strongly reflective ($T_\infty\!\ll\!1$) but non-severing defect
carries only a small boundary entropy relative to a free end.

\section{Unified picture: a one-parameter defect family}
\label{sec:discussion}

\begin{figure}[t]
\centering
\includegraphics[width=0.95\columnwidth]{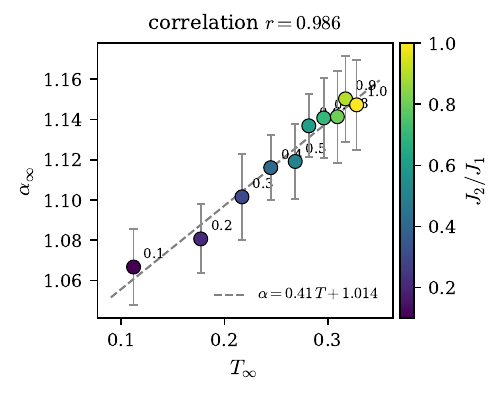}
\caption{The Casimir exponent $\alpha_\infty$ versus the transmission
plateau $T_\infty$, with point color encoding $J_2/J_1$ and error
bars showing the $\alpha_\infty$ extrapolation uncertainty. The two
quantities trace a smooth, nearly linear curve,
$\alpha_\infty \simeq 0.41\,T_\infty + 1.014$. Because both are
monotonic functions of the single tuning parameter $J_2$, they move
together coherently as $J_2$ is varied, with no abrupt features.}
\label{fig:correlation}
\end{figure}

Before presenting the unified picture, we make the logical structure
of the argument explicit, since the case for a continuously varying
conformal defect rests on two distinct legs. That the vacancy is a
\emph{conformal} (scale-invariant) defect at each fixed $J_2$ is
certified by the \emph{form} of the observables, not by their variation
with $J_2$: the clean algebraic decay of $\Delta_b(r)$
[Eq.~\eqref{eq:Db}, Fig.~\ref{fig:alpha}] is the signature of a
scale-invariant Casimir interaction--a relevant defect would flow under
the renormalization group and produce a crossover rather than a single
power law over a decade in $r$--while the saturation of $T(k)$
[Eq.~\eqref{eq:T}, Fig.~\ref{fig:transmission}] to a finite plateau with
$0<T_\infty<1$ shows that the defect neither renormalizes to a severed
chain ($T_\infty\to 0$, two free boundaries) nor heals to a trivial
defect ($T_\infty\to 1$), but instead sits at a marginal, partially
transmissive fixed point; the finite, well-defined $\log g_\infty$
completes this picture. That these defects form a \emph{continuous
family}, in turn, follows from a known property of the Ising
universality class ($c=1/2$): a localized bond defect is exactly
marginal, generating a line of conformal defects parametrized by a
single transmission amplitude, from the trivial defect to two decoupled
free ends~\cite{OshikawaAffleck1996,OshikawaAffleck1997,EislerPeschel2010};
our data show that $J_2$ slides the vacancy continuously along this
line, with $\eta=1/4$ throughout (Sec.~\ref{sec:method}) licensing the
extension to $J_2\neq 0$, where the model is no longer free-fermionic
but the bulk remains Ising. Within this logic the three observables are
not on an equal footing: $T_\infty$ is the family parameter and carries
the claim, whereas $\alpha_\infty$ and $\log g_\infty$ act as
consistency checks that necessarily track it rather than as independent
corroborations. The near-constancy of $\log g_\infty$ in particular is
not a separate axis of the family but exactly what is expected of a
strongly reflective yet non-severing defect, for which the boundary
entropy is small.

The three numbers extracted above--$\alpha_\infty$, $T_\infty$, and
$\log g_\infty$--all vary smoothly and monotonically with the single
tuning parameter $J_2/J_1$. Figure~\ref{fig:correlation} shows
$\alpha_\infty$ versus $T_\infty$; the points trace a smooth, nearly
linear curve, $\alpha_\infty \simeq 0.41\,T_\infty + 1.014$, with
Pearson correlation $r=0.986$. Because both quantities are monotonic
in $J_2$, this correlation is largely a consequence of their common
dependence rather than independent corroboration; the content of
Fig.~\ref{fig:correlation} is that the two observables move together
coherently and without abrupt features as $J_2$ is varied. As $J_2$
increases, the defect becomes more transmissive ($T_\infty\uparrow$)
and the effective Casimir exponent drifts upward ($\alpha\uparrow$).

This coherent behavior is consistent with the following physical picture. The
vacancy in the critical chain $J_1$--$J_2$ is a partially transmissive
conformal defect, and the second-neighbor coupling $J_2$ tunes it
continuously along the one-parameter family of Ising defects discussed
above: small $J_2$
gives a strongly reflective defect with $T_\infty\simeq 0.11$ and
$\alpha\simeq 1.07$, while $J_2/J_1=1.0$ gives a more transparent
defect with $T_\infty\simeq 0.33$ and $\alpha\simeq 1.15$. The small
deviation of $\alpha$ from the universal Casimir value $\alpha=1$
plausibly reflects the subleading correction associated with 
imperfect transmission. The boundary entropy $\log g$ varies little
in this range because $T_\infty$ remains well below unity, so the
defect remains in the strongly reflective regime where boundary
effects are small.

Several features of our results connect with the conformal-defect
literature. Power-law Casimir forces between conformal boundaries are
a generic prediction of CFT boundary
~\cite{AffleckLudwig1991,FriedanKonechny2004}; in particular, the
Casimir energy between two permeable conformal walls is controlled by
their reflection/transmission amplitude~\cite{Bachas2002}, consistent
with our observation that $\alpha$ is close to unity with a prefactor
that tracks $T_\infty$. The finite plateau in $T(k)$ matches the
qualitative behavior expected for non-topological (partially
transmissive) defects in Ising 
CFT~\cite{RoySaleur2022,QuellaRunkelWatts2007}.

We do not provide an explicit analytic mapping between $J_2/J_1$ and
the conformal-defect parameter of, e.g., the Roy-Saleur family. Such
a mapping would require the construction of the lattice defect operator and
projecting onto the relevant scaling fields, which is beyond the
scope of this work. However, our results are tightly constrained:
any candidate defect family must reproduce the smooth and monotonic
relation between $\alpha_\infty$ and $T_\infty$ in
Fig.~\ref{fig:correlation} and the boundary entropy
$\log g_\infty\approx -0.073$, with the values $T_\infty(J_2/J_1)$
reported in Fig.~\ref{fig:transmission}(b).

\section{Conclusions}
\label{sec:conclusion}

We have characterized two static nonmagnetic vacancies in the
critical $J_1$--$J_2$ transverse-field Ising chain by DMRG in open
chains of up to $N=300$ sites, extending our earlier study of the
gapped regime~\cite{SilvaGuimaraesPereira2005} to the quantum
critical line. The vacancy-vacancy binding energy decays
algebraically, $|\Delta_b(r)|\sim r^{-\alpha}$, with an extrapolated
exponent close to unity, $\langle\alpha_\infty\rangle = 1.12\pm 0.03$,
which increases systematically with $J_2/J_1$
[Eq.~\eqref{eq:alpha_fit}]. The transmission of the bulk spin
correlator across a single vacancy saturates to a plateau dependent on $J_2$
 $T_\infty(J_2) \in [0.11,0.33]$. The Affleck-Ludwig boundary
entropy $\log g_\infty \approx -0.073$ is approximately constant
throughout the studied range and lies well above the Ising fixed-BC
value, close to the free-boundary value.

These three observables vary smoothly and monotonically with
$J_2/J_1$, consistent with the vacancy realizing a one-parameter
family of conformal defects, ranging from strongly reflective at
small $J_2$ to more transmissive at $J_2/J_1 \sim 1$. Throughout, the
critical line is located using the bulk order-parameter exponent
$\eta=1/4$, which provides a clean and robust calibration in this
open geometry.

Several extensions of this work are natural. An analytical mapping of
$J_2/J_1$ to a parameter of a known conformal defect family would
turn our numerical relationship between $\alpha_\infty$ and $T_\infty$
into a quantitative prediction. The connection between $\alpha$,
$T_\infty$ and the $g$-theorem~\cite{FriedanKonechny2004} throughout the
family deserves further analytical investigation. The present
setup generalizes naturally to interacting impurities at finite
density and to other quasi-one-dimensional critical magnets.

\begin{acknowledgments}
We acknowledge the support of the Brazilian agencies FAPES, FAPEMIG,
CNPq, and CAPES. A.~R.~Pereira also thanks INCT/CNPq - Spintr\^onica e
Nanoestruturas Magn\'{e}ticas Avan\c{c}adas (INCT-SpinNanoMag) and
Rede Mineira de Nanomagnetismo/FAPEMIG.
\end{acknowledgments}

\appendix

\section{Per-point calibration data and the boundary-entropy extraction}
\label{app:data}

\begin{figure}[t]
\centering
\includegraphics[width=\columnwidth]{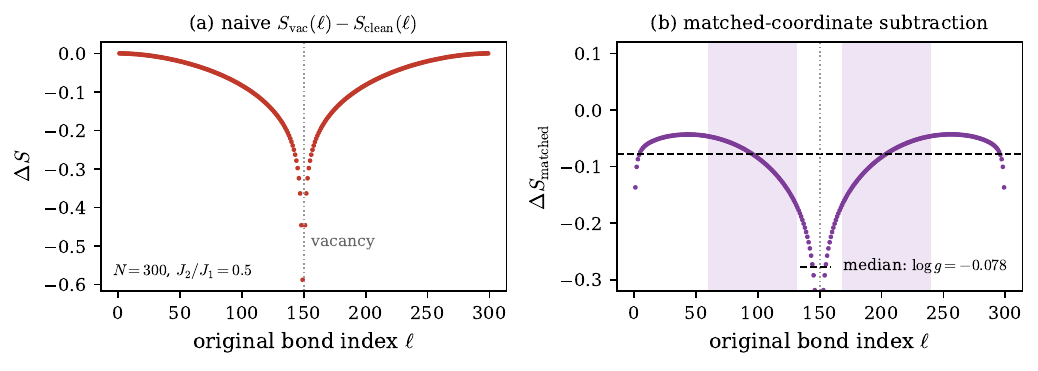}
\caption{Extraction of the boundary entropy, illustrated for
$N=300$, $J_2/J_1=0.5$. (a)~The naive index-matched subtraction
$S_{\rm vac}(\ell)-S_{\rm clean}(\ell)$ produces a deep cusp centered
on the vacancy and no well-defined residual constant, because the
vacancy chain has $N-1$ sites and, for $\ell>v$, a unit shift in the
reduced bond position; the bulk logarithms do not cancel.
(b)~Subtracting the bulk Calabrese--Cardy logarithm at matched
conformal coordinates (effective length $N-1$, reduced position,
$c=1/2$) cancels the bulk contribution; the residual varies slowly
across the bulk window (shaded), reflecting the entanglement
crossover around the defect, and its median gives $\log g$.}
\label{fig:logg_method}
\end{figure}

\begin{table*}[t]
\caption{\label{tab:calib}The 50 calibrated points. $b_c$ from
$\eta=1/4$ tuning; $c_{\rm fit}$ is the central-window (biased) OBC
entanglement estimate; $\alpha$, $T$, and $\log g$ are per-$(N,J_2)$
values (see text). The extrapolated $\alpha_\infty$, $T_\infty$, and
$\log g_\infty$ reported in the main text follow from the $1/N$
extrapolation of the corresponding columns.}
\centering
\scriptsize
\setlength{\tabcolsep}{3pt}
\begin{tabular}{cccccccc@{\qquad}cccccccc}

$J_2/J_1$ & $N$ & $b_c$ & $\eta$ & $c_{\rm fit}$ & $\alpha$ & $T$ & $\log g$  & $J_2/J_1$ & $N$ & $b_c$ & $\eta$ & $c_{\rm fit}$ & $\alpha$ & $T$ & $\log g$  \\

0.1 & 100 & 1.1522 & 0.250 & 0.794 & 1.078 & 0.115 & $-0.0862$ & 0.6 & 100 & 1.8968 & 0.250 & 0.808 & 1.122 & 0.294 & $-0.0811$ \\
 & 150 & 1.1572 & 0.249 & 0.777 & 1.098 & 0.114 & $-0.0839$ &  & 150 & 1.9055 & 0.247 & 0.791 & 1.145 & 0.292 & $-0.0791$ \\
 & 200 & 1.1598 & 0.253 & 0.757 & 1.090 & 0.113 & $-0.0812$ &  & 200 & 1.9105 & 0.254 & 0.764 & 1.143 & 0.287 & $-0.0758$ \\
 & 250 & 1.1610 & 0.251 & 0.757 & 1.075 & 0.113 & $-0.0812$ &  & 250 & 1.9130 & 0.255 & 0.752 & 1.134 & 0.286 & $-0.0746$ \\
 & 300 & 1.1616 & 0.245 & 0.768 & 1.058 & 0.114 & $-0.0824$ &  & 300 & 1.9142 & 0.251 & 0.757 & 1.121 & 0.287 & $-0.0749$ \\[2pt]
0.2 & 100 & 1.3098 & 0.253 & 0.791 & 1.097 & 0.181 & $-0.0841$ & 0.7 & 100 & 2.0365 & 0.250 & 0.810 & 1.126 & 0.308 & $-0.0806$ \\
 & 150 & 1.3154 & 0.251 & 0.775 & 1.116 & 0.179 & $-0.0819$ &  & 150 & 2.0465 & 0.250 & 0.785 & 1.156 & 0.304 & $-0.0777$ \\
 & 200 & 1.3178 & 0.247 & 0.774 & 1.096 & 0.180 & $-0.0815$ &  & 200 & 2.0515 & 0.254 & 0.763 & 1.150 & 0.301 & $-0.0750$ \\
 & 250 & 1.3198 & 0.253 & 0.752 & 1.094 & 0.177 & $-0.0789$ &  & 250 & 2.0540 & 0.253 & 0.756 & 1.138 & 0.300 & $-0.0743$ \\
 & 300 & 1.3204 & 0.246 & 0.767 & 1.076 & 0.179 & $-0.0805$ &  & 300 & 2.0552 & 0.249 & 0.764 & 1.123 & 0.302 & $-0.0750$ \\[2pt]
0.3 & 100 & 1.4612 & 0.248 & 0.806 & 1.097 & 0.226 & $-0.0843$ & 0.8 & 100 & 2.1740 & 0.249 & 0.814 & 1.128 & 0.320 & $-0.0803$ \\
 & 150 & 1.4688 & 0.254 & 0.769 & 1.134 & 0.220 & $-0.0797$ &  & 150 & 2.1852 & 0.251 & 0.783 & 1.163 & 0.315 & $-0.0768$ \\
 & 200 & 1.4712 & 0.248 & 0.775 & 1.109 & 0.222 & $-0.0801$ &  & 200 & 2.1902 & 0.253 & 0.767 & 1.153 & 0.312 & $-0.0748$ \\
 & 250 & 1.4731 & 0.248 & 0.766 & 1.099 & 0.221 & $-0.0791$ &  & 250 & 2.1928 & 0.251 & 0.762 & 1.140 & 0.312 & $-0.0744$ \\
 & 300 & 1.4744 & 0.251 & 0.756 & 1.094 & 0.219 & $-0.0778$ &  & 300 & 2.1940 & 0.245 & 0.775 & 1.123 & 0.315 & $-0.0754$ \\[2pt]
0.4 & 100 & 1.6098 & 0.250 & 0.803 & 1.110 & 0.255 & $-0.0826$ & 0.9 & 100 & 2.3095 & 0.248 & 0.819 & 1.129 & 0.331 & $-0.0802$ \\
 & 150 & 1.6172 & 0.249 & 0.784 & 1.134 & 0.252 & $-0.0802$ &  & 150 & 2.3220 & 0.252 & 0.782 & 1.168 & 0.324 & $-0.0762$ \\
 & 200 & 1.6210 & 0.252 & 0.764 & 1.127 & 0.249 & $-0.0777$ &  & 200 & 2.3270 & 0.251 & 0.772 & 1.155 & 0.323 & $-0.0748$ \\
 & 250 & 1.6229 & 0.251 & 0.759 & 1.114 & 0.249 & $-0.0772$ &  & 250 & 2.3295 & 0.247 & 0.775 & 1.138 & 0.324 & $-0.0751$ \\
 & 300 & 1.6242 & 0.252 & 0.754 & 1.106 & 0.248 & $-0.0764$ &  & 300 & 2.3320 & 0.252 & 0.757 & 1.137 & 0.321 & $-0.0731$ \\[2pt]
0.5 & 100 & 1.7540 & 0.247 & 0.814 & 1.110 & 0.278 & $-0.0827$ & 1.0 & 100 & 2.4450 & 0.252 & 0.810 & 1.140 & 0.337 & $-0.0787$ \\
 & 150 & 1.7628 & 0.248 & 0.787 & 1.141 & 0.274 & $-0.0795$ &  & 150 & 2.4575 & 0.254 & 0.778 & 1.175 & 0.331 & $-0.0753$ \\
 & 200 & 1.7665 & 0.247 & 0.779 & 1.126 & 0.274 & $-0.0783$ &  & 200 & 2.4625 & 0.252 & 0.772 & 1.160 & 0.331 & $-0.0742$ \\
 & 250 & 1.7690 & 0.249 & 0.765 & 1.119 & 0.271 & $-0.0769$ &  & 250 & 2.4650 & 0.246 & 0.777 & 1.141 & 0.333 & $-0.0749$ \\
 & 300 & 1.7702 & 0.246 & 0.770 & 1.107 & 0.272 & $-0.0771$ &  & 300 & 2.4675 & 0.249 & 0.764 & 1.138 & 0.330 & $-0.0733$ \\

\end{tabular}
\end{table*}

Table~\ref{tab:calib} lists the full set of 50 calibrated points
($5$ system sizes $\times$ $10$ values of $J_2/J_1$). For each
$(N,J_2)$ we give the critical field $b_c$ obtained by tuning
$\eta\to1/4$, the resulting bulk correlator exponent $\eta$, the
entanglement central charge $c_{\rm fit}$ from the central-window
Calabrese--Cardy fit, the Casimir exponent $\alpha$ from the window
$r\in[4,32]$, the transmission plateau $T$, and the boundary entropy
$\log g$ from the matched-coordinate subtraction
(Sec.~\ref{sec:method}). We stress that $c_{\rm fit}\simeq0.75$ is the
biased open-boundary estimate discussed in
Sec.~\ref{sec:method} (it reproduces in exact free-fermion data for
the clean chain and is corrected to $c\to1/2$ by the half-chain
estimator); it is tabulated only for completeness and is not used to
locate the critical line.

Figure~\ref{fig:logg_method} illustrates the boundary-entropy
extraction. The naive index-matched subtraction
$S_{\rm vac}(\ell)-S_{\rm clean}(\ell)$ produces a deep cusp centered
on the vacancy and no well-defined residual constant, because the
vacancy chain has $N-1$ sites and, for $\ell>v$, a unit shift in the
reduced bond position; the bulk logarithms do not cancel. Subtracting
the bulk Calabrese--Cardy logarithm at matched conformal coordinates
(effective length $N-1$, reduced position, $c=1/2$) cancels the bulk
contribution; the residual profile varies slowly across the bulk
window (shaded), reflecting the entanglement crossover around the
defect, and its median gives the $\log g$ values tabulated in
Table~\ref{tab:calib}.

\bibliography{biblio}

@article{SilvaGuimaraesPereira2005,
title = {Field-induced nonmagnetic impurities interaction in the quantum Ising chain},
journal = {Solid State Communications},
volume = {134},
number = {5},
pages = {313-317},
year = {2005},
issn = {0038-1098},
doi = {10.1016/j.ssc.2005.02.001},
author = {Ricardo L. Silva and Paulo R.C. Guimarães and Afranio R. Pereira}
}

@article{AffleckLudwig1991,
  author  = {Ian Affleck and Andreas W. W. Ludwig},
  title   = {Universal Noninteger Ground-State Degeneracy in Critical Quantum Systems},
  journal = {Physical Review Letters},
  volume  = {67},
  pages   = {161--164},
  year    = {1991},
  doi     = {10.1103/PhysRevLett.67.161}
}

@article{FriedanKonechny2004,
  author  = {Daniel Friedan and Anatoly Konechny},
  title   = {On the Boundary Entropy of One-Dimensional Quantum Systems at Low Temperature},
  journal = {Physical Review Letters},
  volume  = {93},
  pages   = {030402},
  year    = {2004},
  doi     = {10.1103/PhysRevLett.93.030402}
}

@article{CalabreseCardy2004,
doi = {10.1088/1742-5468/2004/06/P06002},
year = {2004},
month = {jun},
volume = {2004},
number = {06},
pages = {P06002},
author = {Pasquale Calabrese and John Cardy},
title = {Entanglement entropy and quantum field theory},
journal = {Journal of Statistical Mechanics: Theory and Experiment}
}

@article{Calabrese2009,
doi = {10.1088/1751-8121/42/50/500301},
url = {https://doi.org/10.1088/1751-8121/42/50/500301},
year = {2009},
month = {dec},
publisher = {},
volume = {42},
number = {50},
pages = {500301},
author = {Pasquale Calabrese and John Cardy and Benjamin Doyon},
title = {Entanglement entropy in extended quantum systems},
journal = {Journal of Physics A: Mathematical and Theoretical}
}

@article{RoySaleur2022,
  title = {Entanglement Entropy in the Ising Model with Topological Defects},
  author = {Roy, Ananda and Saleur, Hubert},
  journal = {Phys. Rev. Lett.},
  volume = {128},
  issue = {9},
  pages = {090603},
  numpages = {6},
  year = {2022},
  month = {Mar},
  publisher = {American Physical Society},
  doi = {10.1103/PhysRevLett.128.090603},
  url = {https://link.aps.org/doi/10.1103/PhysRevLett.128.090603}
}

@article{OshikawaAffleck1996,
  title = {Defect Lines in the Ising Model and Boundary States on Orbifolds},
  author = {Oshikawa, Masaki and Affleck, Ian},
  journal = {Phys. Rev. Lett.},
  volume = {77},
  issue = {13},
  pages = {2604--2607},
  numpages = {0},
  year = {1996},
  month = {Sep},
  publisher = {American Physical Society},
  doi = {10.1103/PhysRevLett.77.2604},
  url = {https://link.aps.org/doi/10.1103/PhysRevLett.77.2604}
}

@article{OshikawaAffleck1997,
title = {Boundary conformal field theory approach to the critical two-dimensional Ising model with a defect line},
journal = {Nuclear Physics B},
volume = {495},
number = {3},
pages = {533-582},
year = {1997},
issn = {0550-3213},
doi = {10.1016/S0550-3213(97)00219-8},
url = {https://www.sciencedirect.com/science/article/pii/S0550321397002198},
author = {Masaki Oshikawa and Ian Affleck}
}

@article{Bachas2002,
doi = {10.1088/1126-6708/2002/06/027},
url = {https://doi.org/10.1088/1126-6708/2002/06/027},
year = {2002},
month = {jul},
volume = {2002},
number = {06},
pages = {027},
author = {Constantin Bachas and Jan de Boer and Robbert Dijkgraaf and Hirosi Ooguri},
title = {Permeable conformal walls and holography},
journal = {Journal of High Energy Physics}
}

@article{QuellaRunkelWatts2007,
doi = {10.1088/1126-6708/2007/04/095},
url = {https://doi.org/10.1088/1126-6708/2007/04/095},
year = {2007},
month = {apr},
volume = {2007},
number = {04},
pages = {095},
author = {Thomas Quella and Ingo Runkel and Gérard M.T. Watts},
title = {Reflection and transmission for conformal defects},
journal = {Journal of High Energy Physics}
}

@article{EislerPeschel2010,
author = {Viktor Eisler and Ingo Peschel},
title = {Entanglement in fermionic chains with interface defects},
journal = {Annalen der Physik},
volume = {522},
number = {9},
pages = {679-690},
doi = {10.1002/andp.201000055},
url = {https://onlinelibrary.wiley.com/doi/abs/10.1002/andp.201000055},
year = {2010}
}

@article{VidalLatorreRicoKitaev2003,
  title = {Entanglement in Quantum Critical Phenomena},
  author = {Vidal, G. and Latorre, J. I. and Rico, E. and Kitaev, A.},
  journal = {Phys. Rev. Lett.},
  volume = {90},
  issue = {22},
  pages = {227902},
  numpages = {4},
  year = {2003},
  month = {Jun},
  publisher = {American Physical Society},
  doi = {10.1103/PhysRevLett.90.227902},
  url = {https://link.aps.org/doi/10.1103/PhysRevLett.90.227902}
}

@article{White1992,
  title = {Density matrix formulation for quantum renormalization groups},
  author = {White, Steven R.},
  journal = {Phys. Rev. Lett.},
  volume = {69},
  issue = {19},
  pages = {2863--2866},
  numpages = {0},
  year = {1992},
  month = {Nov},
  publisher = {American Physical Society},
  doi = {10.1103/PhysRevLett.69.2863},
  url = {https://link.aps.org/doi/10.1103/PhysRevLett.69.2863}
}

@article{Schollwock2005,
  title = {The density-matrix renormalization group},
  author = {Schollw\"ock, U.},
  journal = {Rev. Mod. Phys.},
  volume = {77},
  issue = {1},
  pages = {259--315},
  numpages = {0},
  year = {2005},
  month = {Apr},
  publisher = {American Physical Society},
  doi = {10.1103/RevModPhys.77.259},
  url = {https://link.aps.org/doi/10.1103/RevModPhys.77.259}
}

@article{Pfeuty1970,
title = {The one-dimensional Ising model with a transverse field},
journal = {Annals of Physics},
volume = {57},
number = {1},
pages = {79-90},
year = {1970},
issn = {0003-4916},
doi = {10.1016/0003-4916(70)90270-8},
url = {https://www.sciencedirect.com/science/article/pii/0003491670902708},
author = {Pierre Pfeuty}
}

@article{Peschel2003,
doi = {10.1088/0305-4470/36/14/101},
url = {https://doi.org/10.1088/0305-4470/36/14/101},
year = {2003},
month = {mar},
volume = {36},
number = {14},
pages = {L205},
author = {Ingo Peschel},
title = {Calculation of reduced density matrices from correlation functions},
journal = {Journal of Physics A: Mathematical and General}
}

@article{PeschelEisler2009,
doi = {10.1088/1751-8113/42/50/504003},
url = {https://doi.org/10.1088/1751-8113/42/50/504003},
year = {2009},
month = {dec},
volume = {42},
number = {50},
pages = {504003},
author = {Peschel, Ingo and Eisler, Viktor},
title = {Reduced density matrices and entanglement entropy in free lattice models},
journal = {Journal of Physics A: Mathematical and Theoretical}
}

@article{ITensor,
	title = {The ITensor Software Library for Tensor Network Calculations},
	pages = {4},
	author = {Fishman, Matthew and White, Steven and Stoudenmire, Edwin Miles},
	journal = {SciPost Phys. Codebases},
	year = {2022},
	publisher = {SciPost},
	doi = {10.21468/SciPostPhysCodeb.4},
	url = {https://scipost.org/10.21468/SciPostPhysCodeb.4}
}

@article{laflorencie20161,
author	= {Nicolas Laflorencie},
title 	= {Quantum entanglement in condensed matter systems},
journal	= {Physics Reports},
volume	= {646},
pages	= {1-59},
year	= {2016},
issn	= {0370-1573},
doi		= {10.1016/j.physrep.2016.06.008},
url		= {https://www.sciencedirect.com/science/article/pii/S0370157316301582}
}

@article{PhysRevLett.99.185301,
title = {Thermodynamic Casimir Effect in $^{4}\mathrm{He}$ Films near ${T}_{\ensuremath{\lambda}}$: Monte Carlo Results},
author = {Hucht, Alfred},
journal = {Phys. Rev. Lett.},
volume = {99},
issue = {18},
pages = {185301},
numpages = {4},
year = {2007},
month = {Nov},
publisher = {American Physical Society},
doi = {10.1103/PhysRevLett.99.185301},
url = {https://link.aps.org/doi/10.1103/PhysRevLett.99.185301}
}

@article{krech1994,
author  = {Krech, M.},
title   = {The Casimir Effect in Critical Systems},
journal = {WORLD SCIENTIFIC eBooks},
year    = {1994},
month   = jul,
doi     = {10.1142/2434}
}

@article{eggertAffleck1992,
  title = {Magnetic impurities in half-integer-spin Heisenberg antiferromagnetic chains},
  author = {Eggert, Sebastian and Affleck, Ian},
  journal = {Phys. Rev. B},
  volume = {46},
  issue = {17},
  pages = {10866--10883},
  numpages = {0},
  year = {1992},
  month = {Nov},
  publisher = {American Physical Society},
  doi = {10.1103/PhysRevB.46.10866}
}

@article{SCHOLLWOCK201196,
author = {Ulrich Schollwöck},
title = {The density-matrix renormalization group in the age of matrix product states},
journal = {Annals of Physics},
volume = {326},
number = {1},
pages = {96-192},
year = {2011},
issn = {0003-4916},
doi = {10.1016/j.aop.2010.09.012}
}

\end{document}